\documentclass[twocolumn,fleqn,usenatbib]{aastex631}
\usepackage[T1]{fontenc}
\usepackage{physics}
\usepackage{tabularx}
\usepackage{graphicx}	
\usepackage{amsmath}	
\usepackage{amssymb}	

\usepackage{newtxtext,newtxmath}

\begin{document}

\title[Non-thermal emission from a conical stratified outflow]{The non-thermal emission following GW170817 is consistent with a conical radially-stratified outflow with initial Lorentz factor $\lesssim10$}


\author[0009-0003-0141-6171]{Gilad Sadeh}
\affiliation{Dept. of Particle Phys. \& Astrophys., Weizmann Institute of Science, Rehovot 76100, Israel}

\author{Eli Waxman}
\affiliation{Dept. of Particle Phys. \& Astrophys., Weizmann Institute of Science, Rehovot 76100, Israel}

\begin{abstract}
We show that the non-thermal radio to X-ray emission following the neutron star merger GW170817 is consistent with synchrotron emission from a collisionless shock driven into the interstellar medium (ISM) by a conical radially stratified outflow observed $\approx0.25$~rad off-axis, with a power-low mass dependence on momentum, $M(>\gamma\beta)\propto(\gamma\beta)^{-4}$, maximum Lorenz factor $\gamma=10$, opening (half-)angle $\approx0.15$~rad, and total energy of $\approx5\times10^{50}$erg. The temporal dependence of the flux during its rising phase is determined by the radial stratification structure, which determines the rate at which outflow energy is deposited in the ISM. This is in contrast with highly relativistic, $\gamma\approx100$, structured jet models, where the angular jet structure determines the time dependence through the gradual "unveiling" by deceleration of larger angular sections of the jet (which are initially "hidden" by relativistic beaming), typically leading to a predicted flux decline after the peak that is faster than observed. Our model predicts a dependence on the observing angle, which is different than that predicted by highly relativistic jet models. Particularly, similar merger events observed closer to the symmetry axis are predicted to show a similarly extended duration of flux increase with time. Our analysis demonstrates that the data do not require a highly relativistic $\gamma\approx100$ component, but the presence of such a component with opening angle $\ll0.15$~rad and energy $\ll5\times10^{50}$~erg cannot be excluded.
\end{abstract}




\section{Introduction}
GW170817 provided the first opportunity for a detailed investigation of the off-axis synchrotron emission driven by a collimated relativistic outflow, with a long duration rich set of observations of the non-thermal radio, optical and X-ray emission \citep{hallinan_radio_2017,troja_x-ray_2017,lyman_optical_2018,troja_thousand_2020,makhathini_panchromatic_2021} and measurement of emission centroid motion \citep{mooley_superluminal_2018,ghirlanda_compact_2019,mooley_optical_2022}.
\begin{itemize}
    \item The observed spectrum is time independent and follows the expected synchrotron spectrum, $F_\nu\propto \nu^{(1-p)/2}$ (where $p$ is the relativistic electrons' spectral index, $dn_e/d\gamma_e \propto \gamma_e^{-p}$), with $p=2.17$ \citep{margutti_binary_2018,troja_thousand_2020,makhathini_panchromatic_2021}, and with no indication for spectral breaks, implying that the synchrotron cooling, self-absorption and peak frequencies are outside the observed frequency range at all observed times.
    \item The light curve shows an initial achromatic rising phase, $F_\nu\propto t^{0.9}$ until $\approx 150$ days \citep{alexander_decline_2018,dobie_turnover_2018,davanzo_evolution_2018}, followed by a steep decline of $F_\nu\propto t^{-1.9}$ up to $\sim 1000$ days  \citep{makhathini_panchromatic_2021,balasubramanian_gw170817_2022,troja_accurate_2022}.
    \item The radio center of light position was measured by the VLBI at 75, 207, and 230 days after the merger \citep{mooley_superluminal_2018,ghirlanda_compact_2019}, complemented by an astrometric measurement of the optical kilonova location eight days after the merger. These measurements indicate a superluminal motion of the flux centroid, implying a $\gamma\gtrsim5$ outflow  \citep{mooley_superluminal_2018,mooley_optical_2022}.
\end{itemize}

The moderate $F_\nu\propto t^{0.9}$ flux rise is inconsistent with that predicted for an ultra-relativistic jet with a uniform distribution of energy per solid angle ("top-hat") observed off-axis \citep{margutti_binary_2018,mooley_mildly_2018,gill_numerical_2019}. Furthermore, the gamma-ray emission associated with GW170817, GRB 170817A, is also inconsistent with such an interpretation \citep{kasliwal_illuminating_2017,granot_lessons_2017, matsumoto_generalized_2019}.
Thus, jets with various angular structures were considered in an attempt to explain the observations \citep{troja_year_2019,fong_optical_2019,lamb_optical_2019,hajela_two_2019,wu_gw170817_2019,ryan_gamma-ray_2020,takahashi_inverse_2020}.
We note that wide-angle outflow models were initially proposed to account for the observations \citep{mooley_mildly_2018,davanzo_evolution_2018,hotokezaka_synchrotron_2018,nakar_implications_2018,troja_outflow_2018,gill_afterglow_2018}, but were set aside in favor of the 'structured jet' models following the detection of the radio centroid motion \citep{mooley_superluminal_2018}, since wide-angle outflow models did not predict the observed fast motion. Furthermore, the post-peak decline predicted by these models is shallower than observed \citep{makhathini_panchromatic_2021}.

In the highly relativistic "structured jet" models, the jet properties (e.g., energy per unit solid angle) vary as a function of the angle with respect to the jet axis. Most papers fitting structured jet models to the observed non-thermal light curve of GW170817 use semi-analytic codes and analytic estimations, such as "afterglowpy", for the emission predicted for given model parameters, combined with an MCMC parameter estimation pipeline \citep{hajela_two_2019,lamb_optical_2019,ghirlanda_compact_2019,wu_gw170817_2019,troja_year_2019, ryan_gamma-ray_2020, troja_thousand_2020,beniamini_afterglow_2020,makhathini_panchromatic_2021,beniamini_robust_2022,ryan_modelling_2023,mcdowell_revisiting_2023,palmese_standard_2024,pellouin_very_2024,morsony_afterglow_2024,wang_jetsimpy_2024,nedora_multi-physics_2024}. Incorporating an accurate treatment of the jet sideways expansion in semi-analytic methods is highly challenging, and it can considerably modify the predictions for the light curve decline phase.
Additionally, these models encounter difficulties in simultaneously reproducing both the light curve and centroid motion without an additional late-time component. Furthermore,
the jet energy inferred in some of these models is typically $\sim10^{52}$~erg, a relatively high value that corresponds to isotropic equivalent energy $\gtrsim10^{54}$~erg \citep{ryan_modelling_2023,wang_jetsimpy_2024,pellouin_very_2024,nedora_multi-physics_2024}, which is inconsistent with short GRB observations \citep{davanzo_complete_2014}.

A more numeric approach to fitting structured jet models to the data was adopted in several papers, based on using numeric 2D hydrodynamic calculations \citep{mooley_superluminal_2018,gill_numerical_2019,gottlieb_detectability_2019}, approximate 1D hydrodynamic calculations \citep{mooley_optical_2022}, and numerically calibrated (to ultra-relativistic structured jet calculations) analytic models \citep{govreen-segal_analytic_2023}. These analyses typically find a jet opening angle of $\theta_j\sim3^\circ-6^\circ$ and a viewing angle of $\theta_\text{v}\sim15^\circ-30^\circ$ with respect to the jet axis. The decline of the flux after its peak time is typically predicted in these models to be faster than observed up to $\sim1000$~days. This is due to the fact that the phase of increasing observed flux extends up to the time when deceleration unveils the full angular extent of the jet to the observer and is then followed by a rapid flux decline that gradually shallows to an asymptotic $F_\nu\propto t^{-p}$ behavior, while the observed decline is shallower than $t^{-p}$ (with $p=2.17$).

A note is in place here regarding the $\gamma$-ray flash associated with GRB 170817A \citep{goldstein_ordinary_2017,savchenko_integral_2017}, which carries an energy that is smaller by $\sim$4 orders of magnitude compared to that of typical short GRBs (sGRB). The low energy is unlikely due to relativistic beaming suppression, i.e., due to observing a highly relativistic $\Gamma\sim100$ jet that produces a typical sGRB along its axis, at a large off-axis angle, $\theta-\theta_j>1/\Gamma$ where $\theta_j$ is the jet opening angle. This is due mainly to the fact that, in this case, the characteristic photon energy is also reduced by a large factor \citep{granot_lessons_2017,kasliwal_illuminating_2017,matsumoto_constraints_2019}. The mechanism producing the $\gamma$-ray flash is not known 
\citep[different explanations suggested, e.g.,][]{murguia-berthier_neutron_2017,ioka_can_2018,nakar__2018,gottlieb_cocoon_2018,lundman_first-principle_2021,beloborodov_relativistic_2020}, and it is thus difficult to derive constraints on the outflow based on its properties. Assuming that the photon spectrum follows a "Comptonized" spectrum up to $\sim1$~MeV, would imply a lower limit of a few to the Lorentz factor of the plasma emitting the radiation \citep{matsumoto_generalized_2019,matsumoto_constraints_2019}.

In this paper, we show that the non-thermal emission is consistent with synchrotron emission from a collisionless shock driven into the ISM by a conical radially stratified outflow observed off-axis (at an angle comparable to the flow opening angle), with a power-low mass dependence on momentum, $M(>\gamma\beta)\propto(\gamma\beta)^{-4}$ and $\gamma\lesssim10$. In this model, the peak in observed flux is obtained when the reverse shock, which is driven into the ejecta and decelerates it due to the interaction with the ISM, crosses through the ejecta (for highly relativistic, $\gamma\approx100$, outflow the reverse shock crosses the ejecta and all the energy is transferred to the ISM at smaller radii, $r\propto \gamma^{-2/3}$, corresponding to much earlier observed times preceding the first radio detection). A shallow decline post-peak is obtained for radially stratified outflows in cases where the reverse shock crosses through the ejecta before the deceleration reveals the full outflow extent to an off-axis observer.

We note that shock breakout through a wide-angle stratified outflow,  $M(>\gamma\beta)\propto(\gamma\beta)^{-4}$, was suggested as a source of the $\gamma$-ray flash \citep{lundman_first-principle_2021,beloborodov_relativistic_2020}. While the structure they consider is similar to that considered in this paper, the outflow mass required in their model to account for the $\gamma$-ray emission, $10^{-7}M_\odot$ at $\gamma>5$, is smaller by $2-3$ orders of magnitude compared to that we infer from the radio emission. 

The paper is organized as follows. In \S~\ref{sec:model}, we present a simple semi-analytic method to calculate the synchrotron emission from a conical stratified outflow. 2D numeric calculations are presented in \S~\ref{sec:numerical}. The accuracy of our semi-analytic calculations is determined in \S~\ref{sec:verification} through a comparison to the numeric results. In all calculations, we assume that electrons are accelerated by the collisionless shock to power-law energy distribution, $dn_e/d\gamma_e\propto \gamma_e^{-p}$ where $\gamma_e$ is the electron Lorenz factor (in the plasma rest frame), and that fractions $\varepsilon_e$ and $\varepsilon_B$ of the post-shock thermal energy are carried by non-thermal electrons and magnetic fields, respectively. In \S~\ref{sec:GW170817}, we show, using both semi-analytic and numeric calculations, that the stratified outflow model may account for the non-thermal emission following GW170817. Our conclusions are summarized in \S~\ref{sec:conclusions}.

\section{A semi-analytic derivation of the flux produced by a conical stratified outflow}
\label{sec:model}
\subsection{Initial ejecta structure}
\label{sec:geometry}

\begin{figure}
    \includegraphics[width=9cm]{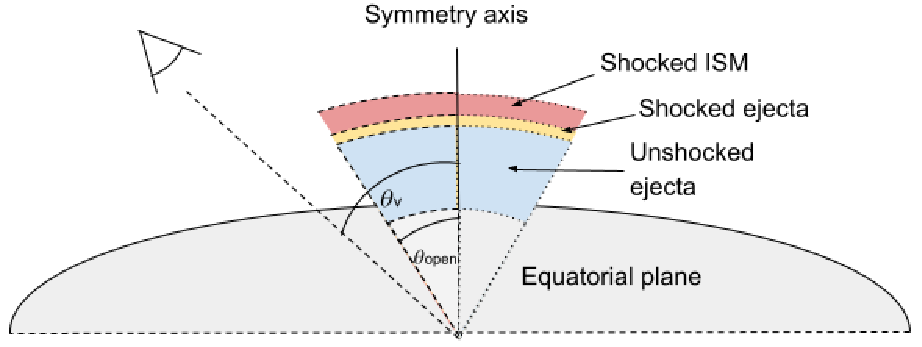}
    \caption{A schematic illustration showing a 2D slice of the 3D conical ejecta (described in \S~\ref{sec:geometry}), and the forward-reverse shock structure that is formed and described in \S~\ref{sec:semi}.
    The symmetry axis corresponds to the azimuthal symmetry of the cone.
    $\theta_\text{open}$ is the initial cone opening angle. $\theta_\text{v}$ is the angle between the symmetry axis and the line of sight, such that $\theta_\text{v}=\pi/2$ corresponds to the equatorial plane.
    }
    \label{fig:geometry}
\end{figure}

We consider a radially stratified double-sided conical ejecta with opening angle  $\theta_\text{open}$ (see Fig. \ref{fig:geometry}), an angle-independent structure, and an initial power-law dependence of mass on momentum,
\begin{equation}
\label{eq:profile}
    M(>\gamma\beta)= M_c\left(\frac{\gamma\beta}{\gamma_c\beta_c}\right)^{-k} \text{for}\quad \gamma_c\beta_c<\gamma\beta,
\end{equation}
where $M_c$ is the equivalent isotropic mass of the conical ejecta.

\subsection{Coordinate systems}
The natural coordinate system for describing the ejecta and its hydrodynamic evolution is a spherical coordinate system $(R,\phi,\theta)$, where $R$ is the distance from the origin, $\phi$ is the azimuthal angle, and $\theta$ is the polar angle measured with respect to the ejecta symmetry axis (see Fig. \ref{fig:geometry}).

To calculate the synchrotron emission, it is convenient to define a coordinate system with a pole aligned with the line of sight, which makes an angle $\theta_{\rm v}$ with the ejecta symmetry axis. We therefore define another spherical coordinate system, $(R,\varphi,\xi)$, where $\varphi$ is the azimuthal angle (chosen such that $\varphi=0$ corresponds to the equatorial plane in the case of $\theta_\text{v}=\pi/2$), $\xi$ is the polar angle, and the pole is aligned with the line of sight. The coordinate transformation is
\begin{equation}
\begin{aligned}
\label{eq:coordinates}\cos\xi&=\cos\theta_\text{v}\cos\theta-\sin{\theta_\text{v}}\sin\theta\sin\phi,\\
        \sin\varphi&=\frac{\cos\theta-\cos\theta_\text{v}\cos\xi}{\sin\xi\sin\theta_\text{v}}.
\end{aligned}
\end{equation}

A cylindrical coordinate system, $(r,\varphi,z)$, such that the $z$-axis is aligned with the line of sight, is useful for the semi-analytic calculation presented below. For this coordinate system, the coordinate transformation is
\begin{equation}
\begin{aligned}
R&=\sqrt{r^2+z^2},\\
    \tan\theta &= \frac{\sqrt{(r\cos{\varphi})^2+(r\sin{\varphi}\cos{\theta_\text{v}}-z\sin{\theta_\text{v}})^2}}{r\sin{\varphi}\sin{\theta_\text{v}}+z\cos{\theta_\text{v}}}.
\end{aligned}
\end{equation}

\subsection{Semi-analytic calculation}
\label{sec:semi}
As the ejecta expands, a forward shock is driven into the ISM, and a reverse shock is driven into the ejecta. In previous papers \citep{sadeh_non-thermal_2023,sadeh_late-time_2024}, we developed a semi-analytic calculation method for modeling the dynamics for the case of a stratified spherical ejecta with a velocity profile given by Eq. (\ref{eq:profile}) propagating into a cold and uniform ISM. 
We have shown that this method provides an accurate approximation for the results of numeric 1D relativistic hydrodynamics calculations \citep{sadeh_non-thermal_2023}. Here, we slightly modify the values of numerically calibrated model parameters to obtain higher accuracy for the higher Lorentz factors considered in this paper.

We approximate the shocked plasma as two uniform layers between the shocks, separated by a contact discontinuity.
In \citet{sadeh_late-time_2024}, we showed that the sideways expansion does not affect the predicted emission significantly up to the time at which the reverse shock crosses the ejecta. This is due to the fact that at any given time, most of the internal energy is deposited into the ISM by the newly decelerated shell of the ejecta, which has not yet undergone significant sideways expansion.

The time at which the reverse shock crosses the ejecta is given, for on-axis observers, by \citep{sadeh_non-thermal_2023}
\begin{equation}
\label{eq:peak_t}
    t_\text{peak}= 550g(\beta_c)\left(\frac{M_{c,-4}}{n_{-2}}\right)^{\frac{1}{3}}\text{days},\quad
   g(\beta_c)=\frac{1.5-\sqrt{0.25+2\beta_c^2}}{\gamma_c^{\frac{1}{3}}\beta_{c}},
\end{equation}
where $n=10^{-2}n_{-2}{\rm cm}^{-3}$ and $M_c=10^{-4}M_{c,-4}M_\odot$. $t_\text{peak}$ is the time at which the flux reaches its peak for an on-axis observer. For an off-axis observer with $\theta_\text{v}>\theta_\text{open}$, the peak time is delayed with respect to $t_\text{peak}$ given by Eq. (\ref{eq:peak_t}) due to the subsequent unveiling of angular sections hidden by relativistic beaming.

Prior to reverse shock crossing, the ejecta structure is well approximated by a conical section of a spherical expanding ejecta.
The emissivity is thus approximated as
\begin{equation}
j'_\nu=
\begin{cases}
j'_\nu \left(e,\rho,\gamma\right) & \text{for}\quad 0<\theta<\theta_\text{open},\\
j'_\nu \left(e,\rho,\gamma\right) & \text{for}\quad \pi-\theta_\text{open}<\theta<\pi,\\
0 & \text{otherwise},
\end{cases}
\end{equation}
where $e,\rho$ and $\gamma$ are the radially dependent plasma proper energy density, proper density, and Lorenz factor, respectively, and the prime ($'$) denotes a quantity measured in the plasma frame.

The emissivity, observed intensity ($I_\nu$), and flux ($F_\nu$) are computed following the methodology outlined in \cite{sadeh_non-thermal_2024} for frequencies above the self-absorption/peak frequency and below the cooling frequency.
Correspondingly, the center of light is calculated by
\begin{equation} R_\text{CoL}=\frac{\frac{1}{D^2}\int\int r^2\sin\varphi I_\nu dr d\varphi }{F_\nu}.
\end{equation}
The coordinates on the sky are chosen such that the symmetry axis is at $\varphi=\pi/2$ (a reflection symmetry axis at $\varphi=\pi/2$), omitting the necessity to consider the $r\cos\varphi$ component.

\section{2D Numerical Calculation}
\label{sec:numerical}
\subsection{Hydrodynamics}
We use the RELDAFNA code \citep{klein_construction_2023} to solve the 2D relativistic hydrodynamics equations. RELDAFNA is an Eulerian code employing the Godunov scheme, incorporating adaptive mesh refinement (AMR) and second-order accuracy in time and space integration. Its efficient parallelization enables high-resolution calculations, even for complex multiscale problems. RELDAFNA's accuracy was confirmed by comparing it with similar codes \citep{zhang_ram_2006, meliani_amrvac_2007} using standard test problems.

In our calculations, we use an equation of state of an ideal fluid with a smoothly varying adiabatic index (between $5/3$ to $4/3$), 
\begin{equation}
\hat{\gamma}=\frac{4+\left(1+1.1\frac{e}{\rho' c^2}\right)^{-1}}{3},
\end{equation}
which is in very good agreement with the equation of state of a mildly relativistic fluid provided by \cite{synge_relativistic_1958}. The $1.1$ factor is introduced for better agreement with the exact solution in the range of $4<e/\rho' c^2<9$ \citep[corresponding to $5<\gamma<10$,][]{sadeh_late-time_2024}.

Our simulations were performed in a 2D cylindrical coordinate system $(r,z)$ assuming axial symmetry and applying reflective boundary conditions at both the symmetry axis ($r=0$) and the symmetry plane ($z=0$). The computational domain boundaries were set at $r=10^{19}$cm and $z=10^{19}$cm.
The initial grid prioritizes resolving the ejecta by placing a higher concentration of cells within it, with a coarser spacing outside. Typically, the simulation begins with $500\times500$ cells, with most cells within the ejecta. AMR dynamically adjusts the resolution throughout the simulation, optimizing cell distribution along both the $r$ and $z$ axes to effectively capture regions of significant variation in pressure, density, or Lorentz factor. We systematically refined the initial spatial grid and temporal steps to assess convergence. We also iteratively increased the number of times the regridding scheme is allowed to multiply the number of cells within a time step. This process was continued until no significant changes were observed in the hydrodynamics and emergent flux.

The initial ejecta structure was chosen as described in \S~\ref{sec:geometry}. The initial radius of a mass shell $M(>\gamma\beta)$ was determined as $r_0=\beta ct_0$, with $t_0=10^6-10^7$s. Behind the ejecta, between $r=\beta_cct_0$ and $r=0.99\beta_cct_0$, we chose a smooth power-law interpolation between $\beta_c$, $\rho_c$ (the density of slowest part the ejecta) and $\beta=0,\rho=10^{-3}\rho_\text{ISM}$ to avoid numeric instabilities due to discontinuities in the density and velocity.
The initial pressure in all of the cells is set to $P=10^{-10}\rho_{\text{ISM}} c^2$ where $\rho_{\text{ISM}}$ is the ISM mass density.

\subsection{Radiation}
The observed flux was calculated following the same procedure described in detail in \citet{sadeh_late-time_2024}. 

For each cell in the 3D grid constructed for the calculation of radiation emission, we calculate the projected center of light along the symmetry axis of the 2D sky plane, $R\sin\xi\sin\varphi$, where $\varphi$ is the azimuthal angle on the sky, see Eq.~(\ref{eq:coordinates}). The center of light is computed by averaging the different projections of each cell weighted by the flux contribution from it ($\Delta F_\nu$),
\begin{equation}R_\text{CoL}=\frac{\Sigma_i^{\#\text{cells}}[R\sin\xi\sin\varphi]_i[\Delta F_\nu]_i}{F_\nu}.
\end{equation}

\section{Semi analytic vs. numerical results}
\label{sec:verification}
\begin{figure*}
    \centering
    \gridline{
        \fig{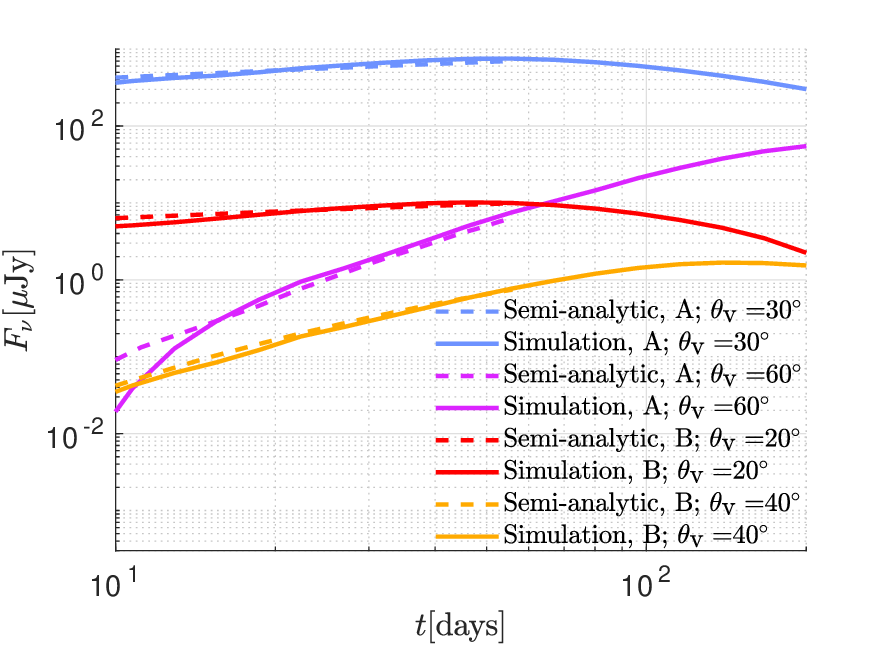}{0.45\textwidth}{(a) Time-dependent flux}
        \fig{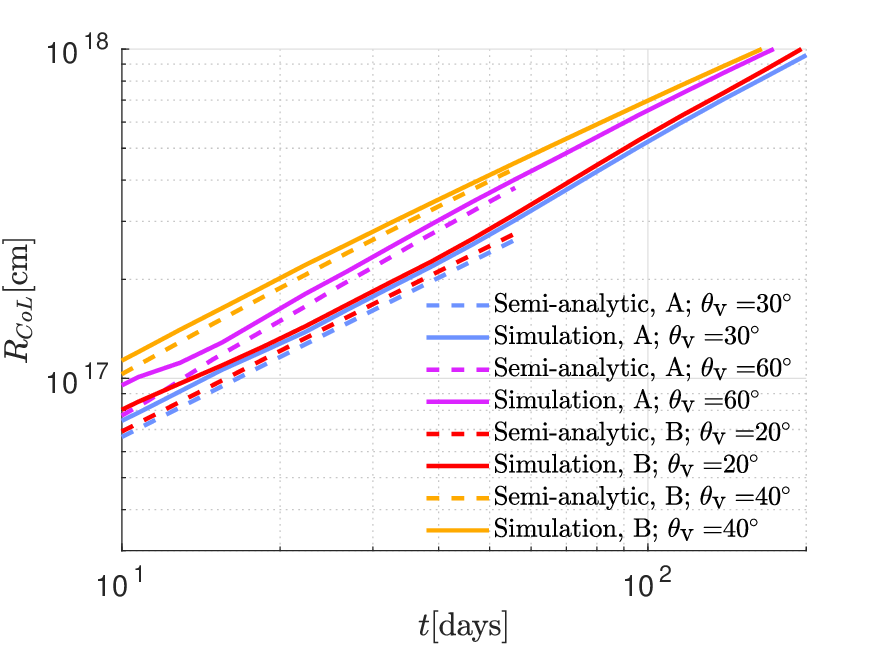}{0.45\textwidth}{(b) Time-dependent light centroid location}
    }
    \caption{A comparison between the time-dependent flux (left panel) and the time-dependent light centroid location (right panel) obtained semi-analytically (solid lines) and numerically (dashed lines), for various viewing angles and two sets of parameter values $\{k,\gamma_c,n,M_c,\varepsilon_e,\varepsilon_B,p,\theta_\text{open}\}=$ A: $\{5,3,10^{-2}\text{cm}^{-3},5.5\times10^{-4}M_\odot,10^{-1},10^{-2},2.2,30^\circ\}$, B: $\{5,3,10^{-3}\text{cm}^{-3},5.5\times10^{-5}M_\odot,10^{-1},10^{-2},2.2,20^\circ\}$. Notice that $\theta_\text{open}$ represents the initial opening angle of the conical ejecta while $\theta_V$ is the viewing angle of the observer. Since the semi-analytic description is valid up to the time at which the reverse shock crosses the ejecta, we compare the two only prior to the peak time scale.
    The flux is shown for a distance of $100$Mpc and a frequency of 6~GHz.}
    \label{fig:comparison}
\end{figure*}

In Fig.~\ref{fig:comparison}, we compare the light curve (at frequencies above the self-absorption and peak frequency and below the cooling frequency), and the time-dependent light centroid location obtained semi-analytically with the results of the 2D numerical calculation. Recall that the semi-analytic description is valid up to the time at which the reverse shock crosses the ejecta, as such, we can compare the two only prior to the peak time scale. Unfortunately, there is no simple analytic expression for the temporal dependence of the rise and decline, which are highly sensitive to the velocity profile, the opening, and the viewing angle.
The agreement is within
$10$'s of percent for relevant values of $\{k,\gamma_c,n,M_c,\varepsilon_e,\varepsilon_B,p,\theta_\text{open}\}$.

\section{GW170817}
\label{sec:GW170817}
\begin{figure}
    \centering
\includegraphics[width=\columnwidth]{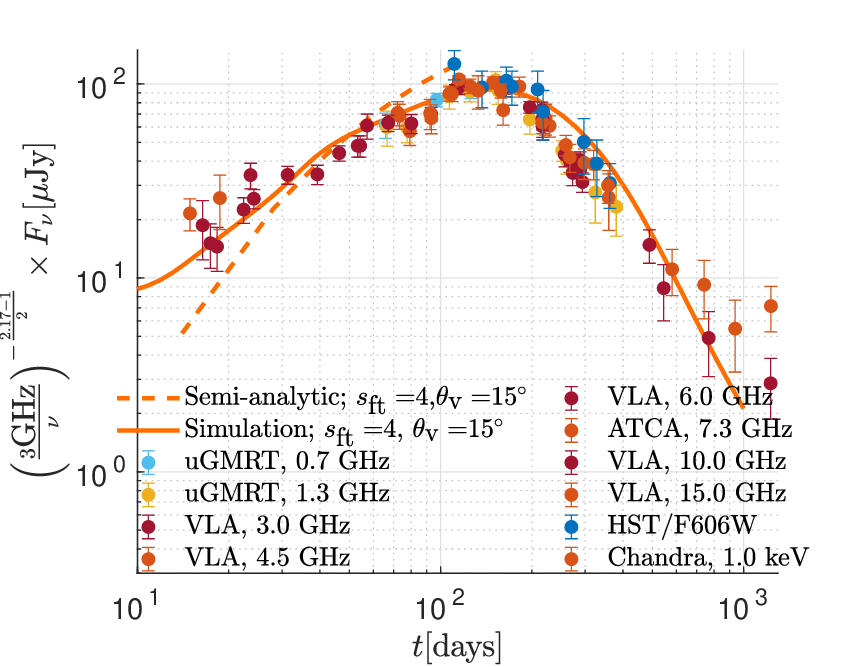}
    \caption{The non-thermal radio to X-ray flux \citep[circles, ][]{makhathini_panchromatic_2021} of the electromagnetic counterpart of GW170817, compared to the results of a 2D numeric calculation of the emission driven by a stratified conical ejecta with $\{k,\gamma_c,n,M_c,\varepsilon_e,\varepsilon_B,p,\theta_\text{open},\theta_\text{v}\}=\{4,5.6,5\times10^{-4}\text{cm}^{-3},5\times10^{-3}M_\odot,0.1,5\times10^{-5},2.17,8^\circ,15^\circ\}$. The model ejecta is truncated beyond $\gamma=10$. The flux is shown for a distance of $41$Mpc.}
    \label{fig:flux_GW170817}
\end{figure}

\begin{figure}
    \centering
\includegraphics[width=\columnwidth]{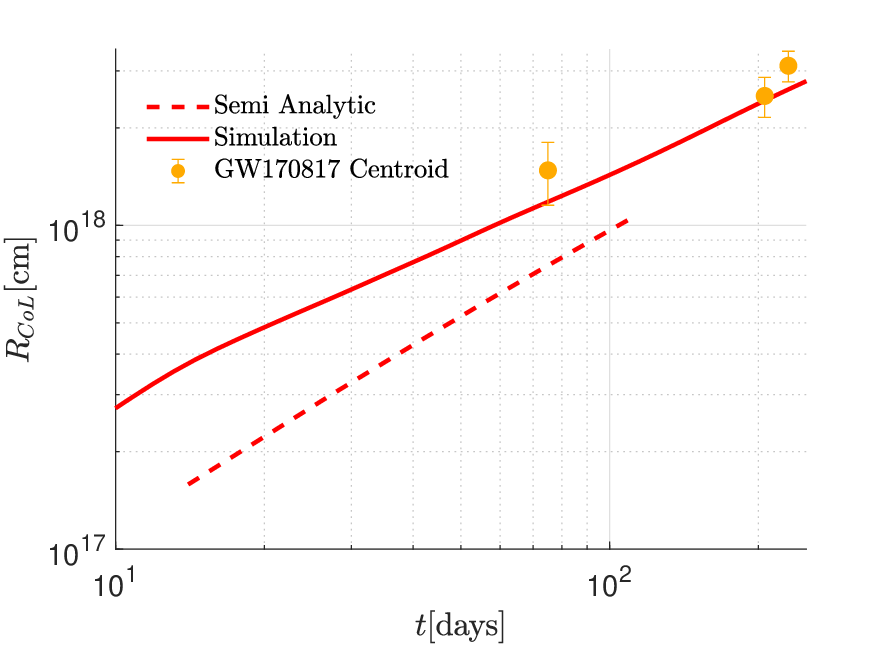}
    The measured motion of the light centroid \citep[circles, ][]{mooley_mildly_2018,mooley_optical_2022} compared to the results of a 2D numeric calculation of the emission driven by a stratified conical ejecta with $\{k,\gamma_c,n,M_c,\varepsilon_e,\varepsilon_B,p,\theta_\text{open},\theta_\text{v}\}=\{4,5.6,5\times10^{-4}\text{cm}^{-3},5\times10^{-3}M_\odot,0.1,5\times10^{-5},2.17,8^\circ,15^\circ\}$. The model ejecta is truncated beyond $\gamma=10$. The centroid motion is shown for a distance of $41$Mpc.
    \label{fig:col_GW170817}
\end{figure}

In \citet{sadeh_synchrotron_2024}, we show that the absence of breaks from the non-thermal power-law spectrum of GW170817 implies that (i) the Lorentz factor of the plasma emitting the observed radiation is $\gamma>2.6$ at $t\sim10$ days and $\gamma<12$ at $t>16$ days, and (ii) $n\cdot\varepsilon_B\lesssim 3\times10^{-7}$cm$^{-3}$. Considering these constraints, we used our semi-analytic method to find parameter values that yield model predictions that are in reasonable agreement with observations. We then used the 2D numeric code to calculate the model predictions more accurately and extend them to times past the validity time of the semi-analytic approximation. We do not attempt to obtain a "best-fit" set of parameter values, as the actual outflow structure is likely more complicated than the simplified structure that we are using for our model calculations, which also includes simplifying assumptions regarding the microphysics of particle acceleration and emission of radiation (e.g., fixed electron and magnetic field energy fractions, independent of shock parameters).
We found that the following choice of parameters yields a good fit to the data: $\{k,\gamma_c,n,M_c,\varepsilon_e,\varepsilon_B,p,\theta_\text{open},\theta_\text{v}\}=\{4,5.6,5\times10^{-4}\text{cm}^{-3},5\times10^{-3}M_\odot,0.1,5\times10^{-5},2.17,8^\circ,15^\circ\}$ \citep{makhathini_panchromatic_2021,mooley_mildly_2018,mooley_optical_2022}, as shown in Figs.~\ref{fig:flux_GW170817}-\ref{fig:col_GW170817}. In the numeric calculation, the ejecta is truncated beyond $\gamma=10$.

\section{Conclusions}
\label{sec:conclusions}
We have shown that the non-thermal radio to X-ray emission following the neutron star merger GW170817 is consistent with synchrotron emission from a collisionless shock driven into the interstellar medium (ISM) by a conical radially stratified outflow observed $\approx0.25$~rad off-axis, with a power-low mass dependence on momentum, $M(>\gamma\beta)\propto(\gamma\beta)^{-4}$ over the range $5.5\lesssim\gamma\lesssim10$, opening (half-)angle $\approx0.15$~rad, and total energy of $\approx5\times10^{50}$erg. A comparison of the measured and model-predicted fluxes and light centroid motion is given in figures~\ref{fig:flux_GW170817} and~\ref{fig:col_GW170817}, for a specific choice of values of model parameters. We did not attempt to obtain a "best-fit" set of parameter values, as the real outflow structure is likely more complicated than the simplified structure that we are using for our model calculations, which also includes simplifying assumptions regarding the microphysics of particle acceleration and emission of radiation (e.g., fixed electron and magnetic field energy fractions, independent of shock parameters). The results of the 2D numeric calculations of the light curve and centroid motion agree with those of our semi-analytic approximation (valid up to the flux peak time, see figure~\ref{fig:comparison}), supporting the validity of both.

The temporal dependence of the flux during its rising phase is determined in the radially stratified outflow model by the radial stratification structure, which determines the rate at which outflow energy is deposited in the ISM. This is in contrast with highly relativistic, $\gamma\approx100$, structured jet models, where the angular jet structure determines the time dependence through the gradual "unveiling" by deceleration of larger angular sections of the jet (which are initially "hidden" by relativistic beaming). In the relativistic jet models, the viewing angle $\theta_\text{v}$ is required to be significantly larger than the opening angle $\theta_\text{open}$,  $\theta_\text{v}>2\theta_\text{open}$ to allow for the observed rising flux phase, and the peak flux is obtained when the deceleration reveals the full angular extent of the jet to the observer, leading to a predicted flux decline after peak that is faster than observed. In the stratified outflow model, the rising flux does not require a large viewing angle, and the post-peak decline is consistent with the observed one.

The radially stratified outflow model predicts a dependence on the observing angle, which is different than that predicted by highly relativistic jet models. Particularly, similar merger events observed at smaller viewing angles are predicted in our model to show a similarly extended duration of flux increase with time. Our analysis demonstrates that the data do not require a highly relativistic $\gamma\approx100$ component, but the presence of such a component with opening angle $\ll0.15$~rad and energy $\ll5\times10^{50}$~erg cannot be excluded, and it may dominate the observed flux when observed on-axis \citep[although slower outflows for gamma-ray bursts afterglows were suggested in][]{dereli-begue_wind_2022}. The rapid deceleration of such a component, compared to the deceleration of the slower $\gamma<10$ outflow, will lead to a light curve that can be discriminated from that of the slower outflow. We note that since flow structures that differ widely ($\gamma\sim100$ relativistic jet and $\gamma<10$ radially stratified outflow) may predict similar non-thermal flux when observed off-axis, 
(with relativistic jets typically predicting a steeper decline),
one cannot accurately infer parameters, e.g., the viewing and opening angles, based on fitting models to the data. This also implies that an accurate determination of $H_0$ \citep{hotokezaka_hubble_2019,govreen-segal_analytic_2023,palmese_standard_2024} is impossible through such parameter inference.

\section*{Acknowledgments}
We thank Y. Klein and N.Wygoda for their authorization to use the RELDAFNA code. We also thank A. Gruzinov for useful discussions.
Eli Waxman's research is partially supported by ISF, GIF and IMOS grants.

\section*{Data Availability}
The data underlying this article will be shared following a reasonable request to the corresponding author.


\bibliography{references}

\bibliographystyle{aasjournal}

\end{document}